# A novel approach for the temperature dependency of saturated signal power in EDFAs

**Cüneyt Berkdemir[*], Sedat Özsoy**

Department of Physics, Faculty of Arts and Sciences, Erciyes University, 38039 Kayseri, Turkey
[*] *Corresponding author: e-mail;* berkdemir@erciyes.edu.tr

**Abstract.**

The temperature dependency of the saturated signal power for the $^4I_{13/2} \to {}^4I_{15/2}$ transition in erbium-doped fiber amplifiers (EDFAs) pumped at 980 nm and 1480 nm-pump wavelengths within a temperature range from − 20 to 60 °C are investigated by a novel approach. For 1480 nm pumping regime, it is seen that the saturated signal power increases more quickly than that of 980 nm pumping regime, with the increasing temperature. The variation in the saturated signal power with temperature is nearly constant at 980 nm pumping regime. In addition, the population inversion with respect to the increasing normalized signal power is examined and it is seen that it is independent of temperature for 980 nm but it strongly depends on temperature for 1480 nm especially at lower normalized signal powers.

## 1. Introduction

Erbium-doped fiber amplifiers (EDFAs) operating in the saturated signal power (SSP) regime have been widely used in optical networks and optical communication systems due to their low crosstalk and low noise figure (Sun *et al*. 1997; Dai *et al*. 1997). The performance analysis of these amplifiers depends on theoretical model to be used, such as the rate or the propagation equations. In some works (Desurvire and Giles 1989; Horowitz *et al*. 1999), the saturation effect of a signal in EDFAs, operating at the signal wavelength of nearly 1530 nm, is only related to pump powers, but the temperature dependency is not included.

In this work, the temperature dependency of the saturated signal power for both 980 nm and 1480 nm pumping wavelengths is investigated at the temperature range from -20 to 60 $^o$C. An expression relating the relative population difference to the signal power normalized by the saturated signal power is also derived.

## 2. Mathematical Model

In general, the stimulated absorption rate, $S_{12}^{sat}$, corresponding the saturated signal power $P_s^{sat}$ is given by $S_{12}^{sat} = \Gamma_{12}\sigma_{12}P_s^{sat}/h\nu_{12}A_{12}$ (Desurvire *et al*. 1994). Here $\sigma_{12}$ is the signal absorption cross section between first and second (intermediate) levels, $h\nu_{12}$ is the photon energy for the $^4I_{15/2} \to {}^4I_{13/2}$ transition, $\Gamma_{12}$ is the mode overlap or the confinement factor at signal wavelength $\lambda_{12}$ and $A_{12}$ is the effective cross-sectional area of the fiber core. 980 nm pumping regime is treated as a three-level amplification system but 1480 nm pumping regime is considered as a two-level amplification system. According to the modified rate equation model (Berkdemir and Özsoy 2004a), for 980 nm pumping regime, the population of third level, N$_3$, is nearly equal to zero and thus the total population N consists of Er$^{3+}$-ions in the remaining levels, i.e. N= N$_1$+ N$_2$.

But, for the 1480 nm pumping regime, the population of the second level is supposed to take form $N_2 = N_{21} + N_{22}$, where the sublevel populations $N_{21}$ and $N_{22}$ are filled with $Er^{3+}$-ions coming from the ground state via the absorbed signal and pump powers, respectively, as shown in Fig.1.

**Figure 1.**

The population distribution of $Er^{3+}$-ions between the two-sublevels within the $^4I_{13/2}$ energy state is governed by the Boltzmann's distribution, for maintaining a constant population at the thermal equilibrium. The thermalization process occurring between sublevels is represented by the nonradiative rates $C_{nr}^u$ (transition to the upper sublevel) and $C_{nr}^\ell$ (transition to the lower sublevel). Thus, under the conditions of overall thermal equilibrium, we have the well-known relation $\beta = N_{22}/N_{21} = C_{nr}^u/C_{nr}^\ell = e^{(-\Delta E_2/k_B T)}$, where $k_B$ is Boltzmann's constant, and $\Delta E_2 = E_{22} - E_{21}$ is the energy difference between the relevant sublevels. Consequently, total population of the two-level system can be written as $N = N_1 + N_{21}(1+\beta)$. In this study, for the two-level amplification system, it is assumed that the $\beta$ parameter takes different values for different temperatures and a value of 200 cm$^{-1}$ is used for the energy interval $\Delta E_2$ between $N_{22}$ and $N_{21}$ at the temperature range from -20 to 60 °C.

Based on the above arguments, the rate equations are modified for three and two level amplification systems, which represent EDFAs pumped by 980 nm and 1480 nm, respectively, and the relations for the relative population difference are given as (Berkdemir and Özsoy 2004a, Berkdemir and Özsoy 2004b):

$$\left(\frac{\Delta N}{N}\right)_{980} = \frac{R_p\tau + S_{12}\tau(1-\eta)-1}{R_p\tau + S_{12}\tau(1+\eta)+1}, \tag{1a}$$

$$\left(\frac{\Delta N}{N}\right)_{1480} = \frac{R_p\tau(1-\beta) + S_{12}\tau(1-\eta)-1}{R_p\tau(1+2\beta) + S_{12}\tau(1+\beta+\eta)+1}. \tag{1b}$$

Note that if β=0, (1b) reduces to (1a). Now, the maximum population difference $\Delta N_{max}$, which is required for the saturated signal power, can be derived. The population difference has its maximum when $S_{12}=0$:

$$\left(\frac{\Delta N_{max}}{N}\right)_{980} = \frac{R_p\tau - 1}{R_p\tau + 1}, \tag{2a}$$

$$\left(\frac{\Delta N_{max}}{N}\right)_{1480} = \frac{R_p\tau(1-\beta)-1}{R_p\tau(1+2\beta)+1}. \tag{2b}$$

Where $R_p$ refers to both pump absorption and pump emission rates ($R_p = \Gamma_p\sigma_p P_p/h\nu_p A_p$; $P_p$ is the pump power), and $\tau$ is the lifetime of second level. If $\Delta N$ equals to $\Delta N_{max}/2$, then $P_s$ becomes the saturated signal power $P_s^{sat}$ and is given by

$$\left(P_s^{sat}\right)_{980} = b_s\left(\frac{(b_p P_p)^2 - 1}{b_p P_p(3\eta - 1) - (3-\eta)}\right), \tag{3a}$$

$$\left(P_s^{sat}\right)_{1480} = b_s\left(\frac{[b_p P_p(1+2\beta)+1][b_p P_p(1-\beta)-1]}{b_p P_p[3\eta(\beta+1)-\beta(\beta+4)-1]+\eta-\beta-3}\right). \tag{3b}$$

Here $\eta$ is the ratio of the signal emission and absorption cross sections, i.e., $\sigma_{21}(\lambda_s)/\sigma_{12}(\lambda_s)$ and obtained by using the McCumber relation (McCumber 1964; Zech 1995),

$b_p = \tau \Gamma_p \sigma_p / h\nu_p A_p$ and $b_s = h\nu_s A_s / \tau \Gamma_s \sigma_{12}$. Now, (3a) and (3b) can be employed to investigate how the saturated signal power changes with the launched pump power in the temperature range from -20 to 60 °C.

By considering the system has three levels for all pumping wavelengths (Iizuka 2002), the relative population difference in terms of the saturated signal power is given as below:

$$\frac{\Delta N}{N} = \frac{\Delta N_{max}}{N} \frac{1}{1 + P_s/P_s^{sat}} \tag{4}$$

In the case of pumping at 980 nm, for a large normalized pump power, the maximum obtainable relative population difference is 1 from (2a), but from (2b) for 1480 nm it changes with the parameter β. Under the assumption that number of levels is different for different pump wavelengths, the following relations are obtained by considering (3) and (4).

$$\left(\frac{\Delta N}{N}\right)_{980} = \left(\frac{\Delta N_{max}}{N}\right)_{980} \cdot \left\{1 + \frac{2(P_p b_p \eta - 1) \times F}{P_p b_p c_p + F \times (1-\eta)(P_p b_p + 1)}\right\}^{-1}, \tag{6}$$

$$\left(\frac{\Delta N}{N}\right)_{1480} = \left(\frac{\Delta N_{max}}{N}\right)_{1480} \cdot \left\{1 + \frac{(2+\beta)\left[P_p b_p (\eta - \beta) - 1\right] \times F}{P_p b_p d_p + F \times (1-\eta)[P_p b_p (1+2\beta) + 1]}\right\}^{-1}, \tag{7}$$

where $c_p = [3\eta - 1] - (3-\eta)$, $d_p = [3\eta(\beta+1) - \beta(\beta+4) - 1] + \eta - \beta - 3$ and $F = P_s/P_s^{sat}$.

## 3. Results and Discussion

For the saturated signal power calculations, we used the values in Table 1 for 980 nm pumping regime and those in Table 2 for 1480 nm. The calculated values of $\eta$ for 980 nm pumping regime at -20, 20 and 60 °C are 1.110, 1.095 and 1.084, respectively. The values of $\eta$ and $\beta$ for 1480 nm at these temperatures are obtained as 1.217, 1.185, 1.161 and 0.33, 0.38, 0.43, respectively.

**Table 1.**

**Table 2.**

In the case of 980 nm pumping regime, Fig.2 shows the saturated signal power as a function of launched pump power at the temperature values of – 20, 20 and 60 °C. This figure shows that the saturated signal power is proportional to not only the pump power but also to the temperature. If the temperature is increased from – 20 to 60 °C, the saturated signal powers increase almost the same amount for all the launched pump powers, and it can be seen that its temperature dependency is almost linear.

Fig.3, which is for 1480 nm pumping regime, shows a different behavior than those of Fig.2, i.e., slops of the curves are not equal and higher for increasing temperatures. This difference is due to β and hence due to the temperature dependency of saturated signal power. The dependence of the saturated signal power on temperature is again almost linear.

**Figure 3.**

The relative population difference versus the normalized signal power is plotted in Fig.4 for both pumping regimes. The uppermost curve is related to (4), which assumes that the system has three levels for all wavelengths, the next curve belongs to 980 nm pumping, for all of the temperatures, and the other curves are connected with 1480 nm pumping for the temperatures of

– 20, 20 and 60 ºC, respectively. It is seen from all the curves that, the relative population difference gradually decreases when the saturated signal power increases. For the pumping at 1480 nm, an increase in the temperature results in a decrease in the relative population difference, especially at lower normalized signal powers, but at 980 nm, it is not influenced from the temperature changes.

**Figure 4.**

## 4. Conclusion

We have used a new approach to describe the temperature dependency of the saturated signal power for EDFAs operated at 980 nm and 1480 nm pumping regimes, and analyzed the results for the temperature range from – 20 to 60 ºC. The expressions for the relative population difference have also been derived in terms of the parameters including the temperature and the saturated signal power for both pumping regimes. In this approach the main point is that the saturated signal power is connected not only to the pumping powers but also to the temperature. It is seen that, for 1480 nm pumping regime the saturated signal power more quickly increases than that of 980 nm pumping regime, with the increasing temperature. On the other hand, at 980 nm pumping regime, the variation of saturated signal power is nearly constant for all the pump powers within the temperature range. Furthermore, the population inversion with respect to normalized signal power is almost independent of temperature for 980 nm but it strongly depends on temperature and increases with decreasing temperature for 1480 nm pumping.

# References

Sun, Y., A. K. Srivastava, J. L. Zyskind, J. W. Sulhoff, C. Wolf, and R. W. Tkach, "Fast power transients in WDM optical networks with cascaded EDFAs," Electron. Lett. **33,** 313-315 (1997).

Dai, H., J. Pan, and C. Lin, "Comparison of optimum EDFA design for in-line amplification in multichannel AM-VSB video lightwave trunking systems," IEEE Photon. Technol. Lett. **9**, 1008 (1997).

Berkdemir, C., and S. Özsoy, "An investigation of the temperature dependency of the relative population inversion and the gain in EDFAs by the modified rate equations," to be submitted to Optics Communications, (2004a).

Desurvire, E., C. R. Giles, "Gain saturation effect in high-speed, multichannel erbium-doped fiber amplifiers at $\lambda$ = 1.53 μm," Journal of Lightwave Technology, **7**, 2095-2104 (1989).

Horowitz, M., C. R. Menyuk, and S. Keren "Modeling the Saturation Induced by Brod-Band Pulsed Amplified in an Erbium-Doped Fiber Amolifier," IEEE Photonics Technology Lett., **11**, 1235-1237 (1999).

Desurvire, E., Erbium-doped fiber amplifiers: principles and applications, (John Wiley & Sons, New York, 1994), Chap.1.

Berkdemir, C., and S. Özsoy, "The performance analysis dependent on temperature for EDFAs pumped at 1480 nm pump wavelength: A theoretical investigation," to be submitted to 30 Europen Conference on Optical Communication, Stockholm, Sweden, (2004b).

McCumber, D. E., "Einstein relations connecting broadband emission and absorption spectra," Physical Review. **136**, (A4), A954-A957 (1964).

Zech, H., "Measurement technique for the quotient of cross sections $\sigma_{21}(\lambda_s)/\sigma_{12}(\lambda_s)$ of erbium-doper fibers," IEEE Photon. Technol. Lett. **7**, 986-988 (1995).

Iizuka, K., Elements of Photonics, Vol. II: For Fiber and Integrates Optics, (Jhon Wiley & Sons, New York, 2002), Chap.13.

Liu, C-K., J-J. Jou, and F-S. Lai, "Second-Order Harmonic Distortion and Optimal Fiber Length in Erbium-Doped Fiber Amplifiers," IEEE Photon. Technol. Lett. **7**, 1412-1414 (1995).

Desurvire, E., J. W. Sulhoff, J. L. Zyskind, and J. R. Simpson, "Study of Spectral Dependence of Gain Saturation and Effect of Inhomogeneous Broadening in Erbium-Doped Aluminosilicate Fiber Aplifiers," IEEE Photon. Technol. Lett. **2**, 653-655 (1990).

## Figure Sublegends

Fig. 1. Three and two-level amplification systems for 980 nm and 1480 nm pumping regime, respectively.

Fig. 2. The dependence of the saturated signal power to the launched pump power in the related temperature values for 980 nm pumping regime.

Fig. 3. The dependence of the saturated signal power to the launched pump power in the related temperature values for 1480 nm pumping regime.

Fig. 4. The comparison of the relative population difference with the normalized signal power in the related temperature values for both pumping regimes. The upper curve is related to Eq.(4) and its lower curve is corresponded to the case of 980 nm pumping regime. The rest curves are related to the case of 1480 nm at -20, 20 and 60 °C, respectively.

**Tables**

**Table 1.** The values of fiber parameter used in our model calculations for 980 nm pumping regime (Liu 1995)

| Parameters | Values | Parameters | Values |
|---|---|---|---|
| $\sigma_{12}$ | $5.25 \times 10^{-25}$ m² | $A_{p,s}$ | $19.6 \times 10^{-12}$ m² |
| $\sigma_{21}$ | $5.75 \times 10^{-25}$ m² | $\Gamma_{p,s}$ | 0.5 |
| $\sigma_p$ | $2.0 \times 10^{-25}$ m² | N | $5.8 \times 10^{24}$ m⁻³ |
| $\tau$ | 11 ms | $\lambda_s$ | 1558 nm |

**Table 2.** The values of fiber parameter used in our model calculations for 1480 nm pumping regime (Desurvire *et al*. 1990)

| Parameters | Values | Parameters | Values |
|---|---|---|---|
| $\sigma_{12}$ | $5.94 \times 10^{-25}$ m² | $A_{p,s}$ | $14.6 \times 10^{-12}$ m² ; $28.3 \times 10^{-12}$ m² |
| $\sigma_{21}$ | $7.04 \times 10^{-25}$ m² | $\Gamma_{p,s}$ | 0.5;  0.3 |
| $\sigma_p$ | $2.0 \times 10^{-25}$ m² | N | $2.2 \times 10^{24}$ m⁻³ |
| $\tau$ | 10 ms | $\lambda_s$ | 1531 nm |